\documentclass{article}
\usepackage{amssymb}
\usepackage{graphicx}
\usepackage{amsmath}

\newtheorem{theorem}{Theorem} [section]

\newtheorem{definition}[theorem]{Definition}
\newtheorem{example}[theorem]{Example}

\newtheorem{lemma}[theorem]{Lemma}

\newtheorem{proposition}[theorem]{Proposition}

\newenvironment{proof}[1][Proof]{\textbf{#1.} }{\ \rule{0.5em}{0.5em}}
\renewcommand{\bibitem}[2]{{#2}\smallskip \newline }
\allowdisplaybreaks[1] 
\begin{document}

\author{\textbf{Luis A. Guardiola}$^{a}$, \textbf{Ana Meca}$^{a}$,
\textbf{Justo Puerto}$^{b,}$
\thanks{Corresponding author. Fax: +34 954622800.\newline
\indent \emph{\ \ E-mail addresses:} {\tt ana.meca@umh.es} (L.
Guardiola
and A. Meca), {\tt puerto@us.es} (J. Puerto). }\vspace*{.5cm} \\
\emph{\small $^{a}$Operations Research Center, Universidad Miguel
Hern\'{a}ndez,}\\
\emph{\small Avda. de la Universidad s/n, Elche, 03202 Alicante,
Spain.}
\\ \emph{\small $^{b}$Facultad de Matem\'aticas, Universidad de Sevilla, 41012
Sevilla, Spain.}}
\title{\textbf{Production-Inventory games: a new class of totally balanced
combinatorial optimization games}\thanks{%
The research of the authors is partially supported by Spanish
Ministry of Science and Technology grants number: BFM2004-0909,
HA2003-0121, MTM2005-09184-C02-02, SEC2002-00112. Authors
acknowledge valuable comments made by the guest Editor and the
referees. }}
\date{}
\maketitle

\begin{abstract}
In this paper we introduce a new class of cooperative games that
arise from production-inventory problems. Several agents have to
cover their demand over a finite time horizon and shortages are
allowed. Each agent has its own unit production, inventory-holding
and backlogging cost. Cooperation among agents is given by sharing
production processes and warehouse facilities: agents in a coalition
produce with \ the cheapest production cost and store with the
cheapest inventory cost. We prove that the resulting cooperative
game is totally balanced and the Owen set reduces to a singleton:
the Owen point. Based on this type of allocation we find a
population monotonic allocation scheme for this class of games.
Finally, we point out the relationship of the Owen point with other
well-known allocation rules such as the nucleolus and the Shapley
value. \medskip

\noindent \textit{JEL classification:} C71

\noindent \textit{2000 AMS Subject classification:} 91A12, 90B05

\noindent \textit{Key words:} Production-inventory games, Totally
balanced combinatorial optimization games, Core-allocations,
Owen-allocations, Monotonicity rules.

\end{abstract}

\newpage

\section{Introduction}

One of the main objectives of management of firms is cost reduction.
In order to achieve this goal, groups of firms tend to form
coalitions to diminish operation costs making dynamic decisions
throughout a finite planning horizon. In tactical planning of
enterprises which produce indivisible goods, operation costs mainly
consist of production, inventory-holding, and backlogging costs.
These coalitions should induce individual and collective cost
reductions; thus, stability is achieved in the process of enterprise
cooperation.

In our framework a coalition allows each of its members of have
access to the technologies owned by the other members of the
coalition. Thus, members of a coalition can use the lowest-cost
technology of the firms in the coalition. Planning is done
throughout a finite time horizon; therefore, at the beginning of
each period, the costs to the members of a coalition, which depend
on the best technology at that point, may change.


The model that represents that situation is the dynamic, discrete,
finite planning horizon production-inventory problem with
backlogging. The objective of any group of firms is to satisfy the
demand for indivisible goods in each period at a minimum cost. This
is a well-known combinatorial optimization problem for which the
algorithm by Wagner \& Whitin provides optimal solutions by dynamic
programming techniques. The optimal solutions of this problem lead
to the best production-inventory policy for the group of firms.
These policies generate an optimal operation cost for the entire
group. The question is what portion of this cost is to be supported
by each firm. Cooperative game theory provides the natural tools for
answering this question.

The analysis of inventory situations is not new. Thus, one can find
in the literature several centralization inventory models approached
from this point of view. The interested reader is referred to
 Eppen (1979), Kohli and
Park (1989), Gerchak and Gupta (1991), Robinson (1993), Hartman and
Dror (1996, 2003, 2005), Hartman et al. (2000), Anupindi et al.
(2001), Muller et al. (2002), Meca et al. (2003) and (2004), Meca
(2006), Minner (2005), Tijs et al. (2005) and Slikker et al. (2006)
among others, for comprehensive literature on this subject. Other
operations research games are studied as well. For a clear and
detailed presentation of operations research games (including
inventory games) we refer to Borm et al. (2001). We are not aware of
any reference in the literature of centralization in inventory
models that  analyzes inventory models for which optimal operation
is only defined implicitly as the optimal solution of a
combinatorial problem (as it is the case in the discrete review
model in this paper). In this regards, our approach makes a step
forward.

The study of cooperative combinatorial optimization games, which are
defined through characteristic functions given as optimal values of
combinatorial optimization problems, is a fruitful topic (see for
instance Shapley and Shubik, 1972, Dubey and Shapley, 1984, Granot,
1986, Tamir, 1992, Deng et al. 1999 and 2000, and Faigle and Kern,
2000). There are characterizations of the total balancedness of
several classes of these games. Inventory games and combinatorial
optimization games are up to date disjoint classes of games. While
in the former class there is always an explicit form for the
characteristic function of each game, the characteristic function of
the games in the latter class it is defined implicitly as the
optimal value of an optimization problem in integer variables.

In this paper we introduce a class of production-inventory games
which combine the characteristics of inventory and combinatorial
optimization games: this class models cooperation on production and
storage of indivisible goods and its characteristic function is
defined implicitly as the optimal value of a combinatorial
optimization problem.   It turns out to be a new class of totally
balanced combinatorial optimization games.

We start by introducing definitions and notations in section
\ref{preli}. In section \ref{PI} we give a complete description of
the production-inventory problem (PI-problem). A natural
variant of this problem is addressed in section \ref%
{PIG}. Several agents, each one facing a PI-problem, decide to
cooperate to reduce costs, and then a production-inventory situation
(PI-situation) arises. Then, for each PI-situation, the
corresponding cooperative game structure, namely
production-inventory game (PI-game), is defined. The main results
(total balancedness and an explicit form for the characteristic
function) are stated in this section. Section \ref{OP} completes the
study of PI-games by showing that the Owen set of a PI-situation
(the set of allocations that are achievable through dual solutions)
shrinks to a singleton: the Owen point. Its explicit form is also
provided, and moreover, it is proved that the Owen point can be
reached through a population monotonic allocation scheme. In
addition, a necessary and sufficient condition for the core of a
PI-game to be a singleton: the Owen point, is presented. We propose
the Owen point as a core-allocation for a PI-game which is easy to
calculate and satisfies good properties. Finally, we point point out
the relationship of the Owen point with some well-known point
solutions in cooperative game theory.

\section{\label{preli}Preliminaries}

Production-inventory games constitute a class of cooperative cost games with
transferable utility (TU games). A TU cost game is a pair $(N,c)$, where $%
N=\left\{ 1,2,...,n\right\} $ is the finite player set and $c:\mathcal{P}%
(N)\rightarrow \mathbb{R}$ the characteristic function satisfying $%
c(\varnothing )=0.$ The subgame related to coalition $S,c_{S},$ is
the restriction of the mapping $c$ to the subcoalitions of $S.$
We denote by lower case letter $s$ the cardinality of set $S$ , i.e. $card(S)=s,$ for all $%
S\subseteq N.$ A cost-sharing vector will be $x\in \mathbb{R}^{n}$
and, for every coalition $S\subseteq N$ we shall write
$x(S):=\sum_{i\in S}x_{i}$ the
cost-sharing to coalition $S$ (where $x(\varnothing )=0).$ 
The core of the game $(N,c)$ consists of those cost-sharing vectors
which allocate the cost of the grand coalition in such a way that
every other coalition pays at most its cost, given by the
characteristic function: $Core(N,c)=\left\{ x\in
\mathbb{R}^{n}\left/ x(N)=c(N)\text{ and }x(S)\leq c(S)\text{ for all }%
S\subset N\right. \right\} .$ In the following, cost-sharing vectors
belonging to the core will be called core-allocations. A cost game $(N,c)$
has a nonempty core if and only if it is balanced (see Bondareva 1963 or
Shapley 1967). It is a totally balanced game if the core of every subgame is
nonempty. Totally balanced games were introduced by Shapley and Shubik in
the study of market games (see Shapley and Shubik, 1969).

A population monotonic allocation scheme (see Sprumont 1990), or
pmas, for the game $(N,c)$ is a collection of vectors $y^{S}\in
\mathbb{R}^{s}$ for
all $S\subseteq N,S\neq \varnothing $ such that $y^{S}(S)=c(S)$ for all $%
S\subseteq N,S\neq \varnothing, $ and $y_{i}^{S}\geq y_{i}^{T}$ for all $%
S\subseteq T\subseteq N$ and $i\in S.$ Note that if $\left(
y^{S}\right) _{\varnothing \neq S\subseteq N}$ is a pmas for
$(N,c),$ then $y^{S}\in Core(N,c_{S})$ for all $S\subseteq N,S\neq
\varnothing .$ Hence, the set of payoff vectors that can be reached
through a pmas can be seen as a refinement of the core. Every cost
game with pmas is totally balanced. However, it is not possible to
get a pmas with any random selection of cost-sharings and there are
totally balanced cost games without pmas. A core-allocation for
$(N,c),$ denoted by $x\in Core(N,c)$, is reached through a pmas if
there exists $\left( y^{S}\right) _{\varnothing \neq S\subseteq N}$
\ for the game $(N,c)$ such that $y_{i}^{N}=x_{i}$ for all $i\in N.$

A game is said to be subadditive when for all disjoint coalitions $S$ and $%
T, $ $c(S\cup T)\leq $ $c(S)+c(T)$ holds. In a subadditive game, it will
always be beneficial for two disjoint coalitions to cooperate and form a
larger coalition. Balanced cost games might not be subadditive but they
always satisfy subadditive inequalities involving the grand coalition.
However, totally balanced cost games are subadditive. A well-known class of
balanced and subadditive games is the class of concave games (see Shapley,
1971).

Finally to simplify the presentation, for a generic problem (P) we
will denote in the following by $val(P)$, $F(P)$ and $O(P)$, the
optimal value, the set of feasible solutions and the set of optimal
solutions of problem (P), respectively.

\section{\label{PI}Production-Inventory problems}

To start with, the basic form of the production-inventory model is
now described (the interested reader is referred to Johnson and
Montgomery 1974, for further details). Demand for a single product
occurs during each of $T$ consecutive time periods that are numbered
1 through $T$. The demand that occurs during a given period can be
satisfied by production during that period, during any earlier
period (as inventory is carried forward) or can be backlogged to be
covered by production at future periods (as backlogged demand is
accumulated). Inventory at period 1 is zero, and inventory at the
end of period $T$ is required to be zero. The model includes
production, inventory and backlogging costs. The objective is to
schedule production so as to satisfy demand at minimum cost.
Formally, a production-inventory problem (PI-problem for short) is a
5-tuple $(T,d,h,b,p)$ where:

\begin{itemize}
\item $T$ is the planning horizon.

\item $d=(d_{1},\ldots ,d_{T})\ge 0,\, d_{t}=$ demand during period $t$, $%
t=1,\ldots ,T$.

\item $h=(h_{1},\ldots ,h_{T})\ge 0,\, h_{t}=$ unit inventory carrying costs
in period $t$, $t=1,\ldots ,T$.

\item $b=(b_{1},\ldots ,b_{T})\ge 0,\, b_{t}=$ unit backlogging carrying
costs in period $t$, $t=1,\ldots ,T$.

\item $p=(p_{1},\ldots ,p_{T})\ge 0,\, p_{t}=$ unit production costs in
period $t$, $t=1,\ldots ,T$.
\end{itemize}

The decision variables of the model are:

\begin{itemize}
\item $q_t=$ production during period $t$.

\item $I_t=$ inventory at hand at the end of period $t$.

\item $E_t=$ backlogged demand at the end of period $t$.
\end{itemize}
The reader may notice that  $I$ and $E$ variables are instrumental
and help to clarify presentation while $q$ is the only actual set of
decision variables. (Once $q$ is known the others can be obtained
solving a system of linear equations.) These decision variables are
required to be in integer quantities. The resolution amounts to the
following mathematical programming formulation:
\begin{eqnarray}
(PI)\quad &\min &\displaystyle\sum_{t=1}^{T}\left(
p_{t}q_{t}+h_{t}I_{t}+b_{t}E_{t}\right)  \notag \\
&\mbox{s.t. }&I_{0}=I_{T}=E_{0}=E_{T}=0,  \label{eq:uno} \\
&&I_{t}-E_{t}=I_{t-1}-E_{t-1}+q_{t}-d_{t},\quad t=1,\ldots ,T,
\label{eq:dos} \\
&&q_{t},\;I_{t},\;E_{t},\mbox{ non-negative, integer},\;t=1\ldots ,T.
\label{eq:tres}
\end{eqnarray}

Constraint (\ref{eq:uno}) ensures initial and final conditions on inventory
and backlogged demand. Matter is conserved, and (\ref{eq:dos}) requires that
the sum of the inventory at the end of a period is the sum of the inventory
at the end of previous period minus consumption. Finally, constraint (\ref%
{eq:tres}) ensures non-negativity and integrality of the variables
in the problem.

We associate with problem $(PI)$ its linear relaxation $(LPI)$. This
problem turns out to be:
\begin{eqnarray}
(LPI)\quad &\min &\displaystyle\sum_{t=1}^{T}\left(
p_{t}q_{t}+h_{t}I_{t}+b_{t}E_{t}\right)  \notag \\
&\mbox{s.t. }&I_{0}=I_{T}=E_{0}=E_{T}=0, \\
&&I_{t}-E_{t}=I_{t-1}-E_{t-1}+q_{t}-d_{t},\;t=1,\ldots ,T, \\
&&q_{t}\geq 0,\;I_{t}\geq 0,\;E_{t}\geq 0,\;t=1\ldots ,T.
\end{eqnarray}%
First of all, we note in passing that if vector $d$ has integer
components all the extreme points of the feasible region of $(LPI)$
are integers. Indeed, the constraint matrix of problem $(LPI)$ is a
$0$, $\pm 1$ valued matrix where each column has at most one $+1$
and at most one $-1$, then it is totally unimodular and the result
is well-known (see e.g. Garfinkel and Nemhauser 1972).

%

Let $(DLPI)$ be the dual of $(LPI)$. This problem is given as
\begin{eqnarray}
(DLPI)\quad  &\max &\sum_{t=1}^{T}y_{t}d_{t}\mbox{   } \label{eq:no7}\\
&\mbox{s.t.}&y_{t}\leq p_{t},\qquad \qquad t=1,\ldots ,T, \label{eq:no8}\\
&&y_{t+1}-y_{t}\leq h_{t},\quad t=1,\ldots ,T-1, \label{eq:no9}\\
&&-y_{t+1}+y_{t}\leq b_{t},\quad t=1,\ldots ,T-1. \label{eq:no10}
\end{eqnarray}

The consequence of the above discussion is that:
\begin{equation*}
val(PI)=val(LPI)=val(DLPI).
\end{equation*}%
Moreover, problems $(PI)$ and $(LPI)$ have integer optimal
solutions.

Let
\begin{eqnarray}
h_{kt} &=&\sum_{r=k}^{t-1}h_{r},\quad \mbox{for any }k<t,t=2,\ldots
,T;h_{k1}=0,k<1, \\
b_{tk} &=&\sum_{r=t}^{k-1}b_{r},\quad \mbox{ for any }k>t,t=1,\ldots
,T-1;b_{Tk}=0,k>T.
\end{eqnarray}%
According to this notation one can check that a feasible solution for $%
(DLPI) $ is given as:
\begin{equation}
y_{t}^{\ast }=\min \Big\{p_{t},\min_{k<t}\{p_{k}+h_{kt}\},\min_{k>t}%
\{p_{k}+b_{tk}\}\Big\},\;t=1,\ldots ,T,  \label{solu}
\end{equation}%
where

\begin{equation*}
p_{k}=\left\{
\begin{array}{cc}
p_{1} & \text{if }k<1, \\
p_{T} & \text{if }k>T.%
\end{array}%
\right.
\end{equation*}

The reader may notice that this solution induces a feasible solution in the
primal problem. In this solution the demand $d_{t}$ that occurs in period $t$
is produced according to the following scheme:
\begin{equation*}
\left\{
\begin{array}{ll}
\mbox{In period }t & \mbox{if }y_{t}^{\ast }=p_{t}, \\
\mbox{In period }k_{h} & \mbox{if }y_{t}^{\ast
}=p_{k_{h}}+h_{k_{h}t},\;k_{h}<t, \\
\mbox{In period }k_{b} & \mbox{if }y_{t}^{\ast
}=p_{k_{b}}+b_{tk_{b}},\;k_{b}>t.%
\end{array}%
\right.
\end{equation*}

\begin{lemma}
\label{maxi}The vector $y^{\ast }$ given in (\ref{solu}) is an
optimal solution of problem $(DLPI)$.
\end{lemma}
\begin{proof}
The reader can easily check that $y^{\ast }$ satisfies all the constraints of problem $%
(DLPI)$. This solution induces a feasible production plan $(q^{\ast
},I^{\ast },E^{\ast })$ in the primal problem $(PI)$ so that $\displaystyle%
\sum_{t=1}^{T}(p_{t}q_{t}^{\ast }+h_{t}I_{t}^{\ast }+b_{t}E_{t}^{\ast
})=\sum_{t=1}^{T}d_{t}y_{t}^{\ast }$. Therefore, both solutions are optimal
in their corresponding problems.
\end{proof}

The reader may notice that if it were assumed that all the unit
costs $h$, $b$ and $p$ are non-negative integers then the solution
$y^{\ast }$ would also be an integer. Moreover, we would like to
have more information about the structure of the optimal solution
set to $(DLPI)$. The next results show that the optimal solution
$y^{\ast }$ given by (\ref{solu}) is the componentwise upper bound
of any optimal solution of $(DLPI)$. Hence, it is called
\emph{maximal optimal solution} for $(DLPI).$

The following technical lemma will be useful to prove the theorem.
In fact, it reveals that vector $y^{\ast }$ is the componentwise
maximum of all feasible solutions. Recall that by $F(DLPI)$ we
denote the feasible solution set for $(DLPI).$

\begin{lemma}
\label{facti}For each $y\in F(DLPI),y_{t}\leq y_{t}^{\ast }\;$for
all $t\in \{1,\ldots ,T\}.$
\end{lemma}
\begin{proof}
Take $y\in F(DLPI).$ Let $p_{k_{h}}+h_{k_{h}t}=\min_{k<t}%
\{p_{k}+h_{kt}\}$ and $p_{k_{b}}+b_{tk_{b}}=\min_{k>t}%
\{p_{k}+b_{tk}\},t=1,\ldots ,T.$

For any $t\in \{1,\ldots ,T\}$ we can distinguish three situations:

(S1) $p_{t}\leq p_{k_{h}}+h_{k_{h}t}$ and $p_{t}\leq
p_{k_{b}}+h_{k_{b}t}.$ Then, $y_{t}^{\ast }=p_{t}$, and by (8)
$y_{t}\leq y_{t}^{\ast }.$

(S2) $p_{k_{h}}+h_{k_{h}t}<p_{t}$ and $%
p_{k_{h}}+h_{k_{h}t}\leq p_{k_{b}}+b_{tk_{b}}.$ Then, $y_{t}^{\ast
}=h_{k_{h}t}+p_{k_{h}}.$ Taking into account that

\begin{equation*}
\begin{array}{l}
y_{k_{h}}\leq p_{k_{h}}, \\
y_{k_{h}+1}\leq h_{k_{h}}+y_{k_{h}}, \\
y_{k_{h}+2}\leq h_{k_{h}+1}+y_{k_{h}+1}%
\leq h_{k_{h}+1}+\left( h_{k_{h}}+p_{k_{h}}\right) =h%
_{k_{h}k_{h}+2}+p_{k_{h}}, \\
.... \\
y_{t}\leq h_{t-1}+y_{t-1}\leq h_{t-1}%
+\left( h_{k_{h}t-1}+p_{k_{h}}\right) =h_{k_{h}t}+p%
_{k_{h}}.%
\end{array}%
\end{equation*}

\noindent we conclude that $y_{t}\leq y_{t}^{\ast }.$

(S3) $p_{k_{b}}+b_{tk_{b}}<p_{t}$ and $%
p_{k_{b}}+b_{tk_{b}}<p_{k_{h}}+h_{k_{h}t}.$ Then,
$y_{t}^{\ast }=b_{tk_{b}}+p_{k_{b}}$; and %
\begin{equation*}
\begin{array}{l}
y_{k_{b}}\leq p_{k_{b}}, \\
y_{k_{b}-1}\leq b_{k_{b}-1}+y_{k_{b}}, \\
y_{k_{b}-2}\leq b_{k_{b}-2}+y_{k_{b}-1}\leq b_{k_{b}-2}+\left(
b_{k_{b}-1}+p_{k_{b}}\right) =b_{k_{b}-2k_{b}}+p_{k_{b}}, \\
\vdots  \\
y_{t}\leq b_{t}+y_{t+1}\leq b_{t}+\left( b_{t+1k_{b}}+p_{k_{b}}\right)
=b_{tk_{b}}+p_{k_{b}}.%
\end{array}%
\end{equation*}
Hence, $y_{t}\leq y_{t}^{\ast }$.
\end{proof}

\begin{theorem}
\label{oss}The optimal solution set of problem $(DLPI)$ is given by%
\begin{equation*}
O(DLPI)=\left\{ y\in F(DLPI)\left/
\begin{array}{cc}
y_{t}\leq y_{t}^{\ast } & \text{if }d_{t}=0 \\
y_{t}=y_{t}^{\ast } & \text{if }d_{t}>0%
\end{array}%
\right. \right\} .
\end{equation*}
\end{theorem}
\begin{proof}
$(\subseteq )$ Take $y\in O(DLPI).$ By lemma \ref{facti}, $%
y_{t}\leq y_{t}^{\ast }\;$for all $t\in \{1,\ldots ,T\}.$
Suppose there exists a period $t\in \{1,...,T\}$ with $d_{t}>0$%
 and $y_{t}<y_{t}^{\ast }.$ Then by lemma \ref{maxi}, $%
val(DLPI)=y_{1}^{\ast }d_{1}+...+y_{T}^{\ast
}d_{T}>y_{1}d_{1}+...+y_{T}d_{T},$ which is a contradiction.

$(\supseteq )$ It is obvious that for all $y\in F(DLPI)$
such that $y_{t}\leq y_{t}^{\ast },$ if $d_{t}=0,$ and $%
y_{t}=y_{t}^{\ast }$ otherwise,%
\begin{equation*}
\sum_{t=1}^{T}d_{t}y_{t}=\sum_{\substack{ t=1 \\ d_{t}>0}}%
^{T}d_{t}y_{t}^{\ast }=\sum_{t=1}^{T}d_{t}y_{t}^{\ast }.
\end{equation*}

Hence by  duality theorem in linear programming, we can conclude
that $y\in O(DLPI).$
\end{proof}

\section{\label{PIG}Production-Inventory games}

Once we have revisited the classical version of the
production-inventory problem, we address a natural variant of this
problem in which several agents, each one facing a PI-problem,
decide to cooperate to reduce costs. Here the cooperation is driven
by sharing technologies in production, inventory carrying and
backlogged demand. Thus, if a group of agents agree on cooperation
then at each period they will produce and pay inventory carrying and
backlogged demand at the cheapest costs among the members of the
coalition. Cooperation in holding and production costs is usual and
has appeared already in literature. Our mode of cooperation in
backlogging is also natural: once a coalition is formed, all its
members pay compensation to customers for delayed delivering
(backlogging cost) of their demands according to the cheapest cost
among the members in the coalition. In some regards, larger
coalitions are stronger and ``squeeze'' a bit more their clients.
Formally, a production-inventory situation (PI-situation) is a
5-tuple $(N,D,H,B,P)$ where $N$ is the set of players $N=\{1,\ldots
,n\}$ and for each player $i$ $(T,d^{i},h^{i},b^{i},p^{i})$ is a
PI-problem.

The reader may notice that
\begin{equation*}
D=[d^1,\ldots,d^n]^{\prime},\quad H=[h^1,\ldots,h^n]^{\prime},\quad
B=[b^1,\ldots,b^n]^{\prime}, \quad P=[p^1,\ldots,p^n]^{\prime};
\end{equation*}
and
\begin{equation*}
d^i=[d_1^i,\ldots,d_T^i]^{\prime},\quad
h^i=[h_1^i,\ldots,h_T^i]^{\prime},\quad
b^i=[b_1^i,\ldots,b_T^i]^{\prime},\quad p^i=[p_1^i,\ldots,p_T^i]^{\prime}.
\end{equation*}

Note that we can associate with each PI-situation $(N,D,H,B,P)$ a cost
TU-game $(N,c)$ with characteristic function $c$ defined as follows: $%
c(\varnothing)=0$ and for any $S\subseteq N,c(S)=val(LPI(S))$, where
\begin{eqnarray*}
(LPI(S))\quad &\min
&\sum_{t=1}^{T}(p_{t}^{S}q_{t}+h_{t}^{S}I_{t}+b_{t}^{S}E_{t}) \\
&\mbox{s.t.}&I_{0}=I_{T}=E_{0}=E_{T}=0 \\
&&I_{t}-E_{t}=I_{t-1}-E_{t-1}+q_{t}-d_{t}^{S},\quad t=1,\ldots ,T, \\
&&q_{t}\geq 0,\;I_{t}\geq 0,\;E_{t}\geq 0,\;\qquad \qquad \quad t=1,\ldots
,T;
\end{eqnarray*}%
with%
\begin{equation*}
p_{t}^{S}=\min_{i\in S}\{p_{t}^{i}\},\;h_{t}^{S}=\min_{i\in
S}\{h_{t}^{i}\},\;b_{t}^{S}=\min_{i\in
S}\{b_{t}^{i}\},\;d_{t}^{S}=\sum_{i\in S}d_{t}^{i}.
\end{equation*}

Every cost TU-game defined in this way is what we call a
production-inventory game (PI-game).

The main result in this section states that every PI-game is totally
balanced, hence subadditive.

\begin{theorem}
\label{totalbal} Let $(N,D,H,B,P)$ be a PI-situation with $D$ being an
integer matrix. The corresponding PI-game $(N,c)$ is totally balanced.
\end{theorem}
\begin{proof}
For any $S\subseteq N$ consider the reduced game $(S,c_{S})$.

Let $y^{\ast }$ be an optimal solution of $(DLPI(S))$. Define
$u^{\ast }=(y^{\ast \prime }d^{i})_{i\in S}\in \mathbb{R}^{s}$. It
is clear that:
\begin{equation*}
\sum_{i\in S}u_{i}^{\ast }=\sum_{t=1}^{T}y_{t}^{\ast
}d_{t}^{S}=val(DLPI(S))=val(LPI(S))=c_{S}(S).
\end{equation*}%
Moreover, for any $R\subseteq S,y^{\ast }$ is a feasible solution for $%
(DLPI(R))$ since the constraint matrix does not depend on $S$ and
the right-hand-side of $(DLPI(S))$ is componentwise smaller than or
equal to the one of $(DLPI(R))$. Hence,
\begin{equation*}
c_{S}(R)=val(DLPI(R))=\displaystyle\hspace*{-2mm}\max_{y\in {\tiny F(DLPI(R))%
}}\sum_{t=1}^{T}d_{t}^{R}y_{t}\geq \sum_{t=1}^{T}d_{t}^{R}y_{t}^{\ast
}=\sum_{i\in R}y^{\ast \prime }d^{i}=\sum_{i\in R}u_{i}^{\ast }.
\end{equation*}%
Therefore, $Core(S,c_{S})\neq \varnothing .$
\end{proof}

We note in passing that although PI-games are totally balanced games, in
general these games are not concave (see example \ref{cf}).

Recently, Deng et al. (1999) and (2000) have studied some families of
combinatorial optimization games, namely packing and covering games, for
which they prove total balancedness. Here, we have presented a different
class exhibiting the same property. The core of the above mentioned classes
of combinatorial optimization games coincides with the set of dual
solutions, i.e. optimal solutions of the dual problem to the one that
defines the characteristic function. However, the same property does not
hold in our class of PI-games as will be shown in the next examples. In
order to study the core set within the class of PI-games we introduce the so
called \textit{Owen set}: the set of allocations that are achievable through
dual solutions (see Gellekom et al., 2000). Formally, the Owen set of $%
(N,D,H,B,P)$ is defined by%
\begin{equation}
Owen(N,D,H,B,P)=\{(y^{\prime }d^{i})_{i\in N}:y\in O(DLPI(N))\}.
\label{owenset}
\end{equation}%
As a straightforward consequence of Theorem \ref{totalbal}, we obtain that

\begin{equation*}
Owen(N,D,H,B,P)\subseteq Core(N,c).
\end{equation*}

Our next example shows that in general there are core-allocations not
achieved with dual solutions.

\begin{example}
Consider the following PI-situation with two periods and two players, namely
$P1$ and $P2$:

$$
\begin{tabular}{|c|c|c||c|c||c|c||c|c|}
\hline
& \multicolumn{2}{|c||}{Demand} & \multicolumn{2}{|c||}{Production} &
\multicolumn{2}{|c||}{Inventory} & \multicolumn{2}{|c|}{Backlogging} \\
\hline
P1 & 10 & 10 & 2 & 2 & 1 & 2 & 1 & 2 \\ \hline
P2 & 8 & 12 & 1 & 1 & 2 & 4 & 2 & 2 \\ \hline
\end{tabular}%
$$

The data above gives rise to the game with characteristic function in the
following table:

\begin{equation*}
\begin{tabular}{|c|c|c||c|c||c|c||c|c||c||}
\hline
& $d_1^S$ & $d_2^S$ & $p_1^S$ & $p_2^S$ & $h_1^S$ & $h_2^S$ & $b_1^S$ & $%
b_2^S$ & $c$ \\ \hline
$\{1\}$ & 10 & 10 & 2 & 2 & 1 & 2 & 1 & 2 & 40 \\ \hline
$\{2\}$ & 8 & 12 & 1 & 1 & 2 & 4 & 2 & 2 & 20 \\ \hline
$\{1,2\}$ & 18 & 22 & 1 & 1 & 1 & 2 & 1 & 2 & 40 \\ \hline
\end{tabular}%
\end{equation*}

The unique optimal solution of $(DLPI(\{1,2\}))$ is $y^{\ast
\prime}=(1,1)$.
This gives $Owen(N,D,H,B,P)=\{(20,20)\}$. However, $Core(N,c)=%
\{(x_{1},x_{2}):0\leq x_{1}\leq 40,0\leq x_{2}\leq 20,x_{1}+x_{2}=40\}.$
\end{example}

In order to get the Owen set one has to solve three optimization problems, $%
(DLPI(\{1,2\})),(DLPI(\{1\}))$ and $(DLPI(\{2\}))$. In general, for a $n$%
-player game $2^{n}-1$ optimization problems would have had to be solved.

For the sake of simplicity, we obtain an explicit form for the
characteristic function of PI-games.

Consider $(DLPI(S))$ the dual of $(LPI(S)).$ Recall that the problem
$(DLPI(S)$ is defined as $(DLPI)$ (see
\eqref{eq:no7}-\eqref{eq:no10}) where cost coefficients and
right-hand side vector are replaced by $p^S$, $h^S$ and $b^S$, and
$d^S$, respectively.

%
Let us denote%
\begin{eqnarray}
h_{kt}^{S} &=&\sum_{r=k}^{t-1}h_{r}^{S},\quad \mbox{for any }k<t,t=2,\ldots
,T;h_{k1}^{S}=0,k<1, \\
b_{tk}^{S} &=&\sum_{r=t}^{k-1}b_{r}^{S},\quad \mbox{ for any }%
k>t,\;t=1,\ldots ,T-1;b_{Tk}^{S}=0,k>T.
\end{eqnarray}%
Similar arguments to the ones given in Section \ref{PI} show that for all $%
S\subseteq N,S\neq \varnothing ,$ the optimal solution set of problem $%
(DLPI(S))$ is

\begin{equation*}
O(DLPI(S))=\left\{ y(S)\in F(DLPI(S))\left/
\begin{array}{cc}
y_{t}(S)\leq y_{t}^{\ast }(S) & \text{if }d_{t}^{S}=0 \\
y_{t}(S)=y_{t}^{\ast }(S) & \text{if }d_{t}^{S}>0%
\end{array}%
\right. \right\} ,
\end{equation*}%
where $y^{\ast }(S),$ the maximal optimal solution, is now given by

\begin{equation}
y_{t}^{\ast }(S)=\min \Big\{p_{t}^{S},\min_{k<t}\{p_{k}^{S}+h_{kt}^{S}\},%
\min_{k>t}\{p_{k}^{S}+b_{tk}^{S}\}\Big\},\;t=1,\ldots ,T,
\end{equation}%
with

\begin{equation*}
p_{k}^{S}=\left\{
\begin{array}{cc}
p_{1}^{S} & \text{if }k<1, \\
p_{T}^{S} & \text{if }k>T.%
\end{array}%
\right.
\end{equation*}

Let
\begin{equation*}
\begin{array}{ccc}
\displaystyle p_{k_{h}}^{S}+h_{k_{h}t}^{S}=\min_{k<t}\{p_{k}^{S}+h_{kt}^{S}%
\}, & \displaystyle p_{k_{b}}^{S}+b_{tk_{b}}^{S}=\min_{k>t}%
\{p_{k}^{S}+b_{tk}^{S}\}, & t=1,\ldots ,T,%
\end{array}%
\end{equation*}%
and

\begin{equation*}
\begin{array}{l}
\displaystyle H_{1}(S)=\left\{ t\in \{1,...,T\}\left/ p_{t}^{S}\leq
p_{k_{h}}^{S}+h_{k_{h}t}^{S}\text{ and }p_{t}^{S}\leq
p_{k_{b}}^{S}+b_{tk_{b}}^{S}\right. \right\} , \\
\mbox{} \\
\displaystyle H_{2}(S)=\left\{ t\in \{2,...,T\}\left/
p_{k_{h}}^{S}+h_{k_{h}t}^{S}<p_{t}^{S}\text{ and }%
p_{k_{h}}^{S}+h_{k_{h}t}^{S}\leq p_{k_{b}}^{S}+b_{tk_{b}}^{S}\right.
\right\} , \\
\mbox{} \\
\displaystyle H_{3}(S)=\left\{ t\in \{1,...,T-1\}\left/
p_{k_{b}}^{S}+b_{tk_{b}}^{S}<p_{t}^{S}\text{ and }%
p_{k_{b}}^{S}+b_{tk_{b}}^{S}<p_{k_{h}}^{S}+h_{k_{h}t}^{S}\right. \right\} .%
\end{array}%
\end{equation*}

Remark that $\left\{ H_{1}(S),H_{2}(S),H_{3}(S)\right\} $ is a
partition of the planning horizon set $\left\{ 1,\ldots ,T\right\}
.$ According to this notation, it can be checked that an explicit
form for the characteristic function of a PI-game is given by
(\ref{chafu}).

\begin{proposition}
Let $(N,D,H,B,P)$ be a PI-situation with $D$ being an integer matrix, and $%
(N,c)$ the corresponding PI-game. Then, for each $S\subseteq N,S\neq
\varnothing $

\begin{equation}
c(S)=\sum_{t\in H_{1}(S)}p_{t}^{S}d_{t}^{S}+\sum_{t\in
H_{2}(S)}d_{t}^{S}\left( p_{k_{h}}^{S}+h_{k_{h}t}^{S}\right) +\sum_{t\in
H_{3}(S)}d_{t}^{S}\left( p_{k_{b}}^{S}+b_{tk_{b}}^{S}\right) .  \label{chafu}
\end{equation}
\end{proposition}
\begin{proof}
Taking into account that for any $S\subseteq N,c(S)=val(DLPI(S))$ and $%
y^{\ast }(S)\in O(DLPI(S)),$ we obtain
$c(S)=\sum_{t=1}^{T}y_{t}^{\ast }(S)d_{t}^{S}.$ Hence, (\ref{chafu})
holds.
\end{proof}

The explicit form above turns out to be expectable. It means that if
we are in a period in which production costs are less than
production-inventory carrying and production-backlogging carrying
costs $(t\in H_{1}(S))$, we should produce all the demand in that
period. On the contrary, if the period is such that production and
production-backlogging carrying costs are greater than
production-inventory carrying ones $(t\in H_{2}(S))$, all the demand
should have been produced in the earlier period with minimum cost
$(k_{h})$ and inventory carried until that period $(t)$. Finally, in
a period where production and production-inventory carrying costs
are greater than production-backlogging carrying ones $(t\in
H_{3}(S)),$ all the demand should be produced in the future period
with minimum cost $(k_{b})$ and backlogging carried until that
period $(t).$

The reader may notice that if the matrices $H$, $B$ and $P$ are in integer
values then the characteristic function of these games is also integer.

The next example illustrates all the results obtained in Section 4.

\begin{example}
\label{cf}Consider the following PI-situation with three periods and three
players:

\begin{equation*}
\begin{tabular}{|c|c|c|c||c|c|c||c|c|c||c|c|c|}
\hline
& \multicolumn{3}{|c||}{Demand} & \multicolumn{3}{|c||}{Production} &
\multicolumn{3}{|c||}{Inventory} & \multicolumn{3}{|c|}{Backlogging} \\
\hline
P1 & 0 & 2 & 5 & 3 & 1 & 2 & 2 & 0 & 1 & 1 & 2 & 1 \\ \hline
P2 & 0 & 8 & 6 & 4 & 2 & 3 & 2 & 0 & 1 & 1 & 3 & 1 \\ \hline
P3 & 0 & 6 & 2 & 3 & 1 & 2 & 2 & 0 & 1 & 1 & 2 & 1 \\ \hline
\end{tabular}%
\end{equation*}

It can be easily checked that for all $S\subseteq
N,H_{1}(S)=\{2\},H_{2}(S)=\{3\},H_{3}(S)=\{1\},$ and $%
c(S)=p_{2}^{S}d_{2}^{S}+d_{3}^{S}(h_{23}^{S}+p_{2}^{S})+d_{1}^{S}(b_{21}^{S}+p_{2}^{S}).
$ Hence, the data above gives rise to the game with characteristic
function in the following table:
\begin{equation*}
\begin{array}{|c|c|c|c||c|c|c||c|c|c||c|c|c||c||}
\hline
& d_{1}^{_{S}} & d_{2}^{_{S}} & d_{3}^{_{S}} & p_{1}^{_{S}} & p_{2}^{_{S}} &
p_{3}^{_{S}} & h_{1}^{_{S}} & h_{2}^{_{S}} & h_{3}^{_{S}} & b_{1}^{_{S}} &
b_{2}^{_{S}} & b_{3}^{_{S}} & c \\ \hline
\{1\} & 0 & 2 & 5 & 3 & 1 & 2 & 2 & 0 & 1 & 1 & 2 & 1 & 7 \\ \hline
\{2\} & 0 & 8 & 6 & 4 & 2 & 3 & 2 & 0 & 1 & 1 & 3 & 1 & 28 \\ \hline
\{3\} & 0 & 6 & 2 & 3 & 1 & 2 & 2 & 0 & 1 & 1 & 2 & 1 & 8 \\ \hline
\{1,2\} & 0 & 10 & 11 & 3 & 1 & 2 & 2 & 0 & 1 & 1 & 2 & 1 & 21 \\ \hline
\{1,3\} & 0 & 8 & 7 & 3 & 1 & 2 & 2 & 0 & 1 & 1 & 2 & 1 & 15 \\ \hline
\{2,3\} & 0 & 14 & 8 & 3 & 1 & 2 & 2 & 0 & 1 & 1 & 2 & 1 & 22 \\ \hline
\{1,2,3\} & 0 & 16 & 13 & 3 & 1 & 2 & 2 & 0 & 1 & 1 & 2 & 1 & 29 \\ \hline
\end{array}%
\end{equation*}

The optimal solution set of $(DLPI(\{1,2,3\}))$ is $\left\{
(y_{1},1,1)\left/ -1\leq y_{1}\leq 2\right. \right\} .$ However, the
Owen
set and the core reduces to a singleton: $Owen(N,D,H,B,P)=Core(N,c)=%
\{(7,14,8)\}.$ Moreover, the game above is not concave since $%
c(\{1,2\})-c(\{2\})=-7<c(N)-c(\{2,3\})=7.$
\end{example}

Note that even though there are infinite optimal solutions of the
problem $(DLPI(N))$, the Owen set of the PI-situation above
reduces to a singleton again. We wonder if every PI-situation
exhibits this property. Next section gives an affirmative answer.

\section{\label{OP}Owen point}

It is well-known that the larger the player set, the more difficult
it is to determine the core of a cost TU-game (i.e. the number of
inequalities involved increases rapidly). The goal of this section
is to find a core-allocation for a PI-game which is easy to
calculate and satisfies good properties.

The first result reveals that the Owen set of a PI-situation shrinks to a
singleton and provides an explicit form to compute it.

\begin{theorem}
Let $(N,D,H,B,P)$ be a PI-situation with $D$ being an integer matrix. Then, $%
Owen(N,D,H,B,P)=\{(o_{1},...,o_{n})\}$ where, for each $i\in N,$
\begin{equation}
o_{i}=\sum_{t\in H_{1}(N)}p_{t}^{N}d_{t}^{i}+\sum_{t\in
H_{2}(N)}d_{t}^{i}\left( p_{k_{h}}^{N}+h_{k_{h}t}^{N}\right) +\sum_{t\in
H_{3}(N)}d_{t}^{i}\left( p_{k_{b}}^{N}+b_{tk_{b}}^{N}\right) .
\label{owenpoint}
\end{equation}
\end{theorem}
\begin{proof}
Take $y^{\ast }(N),y(N)\in O(DLPI(N)),y^{\ast }(N)\neq y(N).$
Proving that 
 $\sum_{t=1}^{T}y_{t}(N)d_{t}^{i}=\sum_{t=1}^{T}y_{t}^{\ast
}(N)d_{t}^{i}$, it follows $o_{i}:=\sum_{t=1}^{T}y_{t}^{\ast
}(N)d_{t}^{i}$ which proves (\ref{owenpoint}).

Taking into account that $d_{t}^{N}=0$ if and only if $d_{t}^{i}=0$ for all $%
i\in N,$ and $y_{t}(N)=y_{t}^{\ast }(N)$ if $d_{t}^{N}>0,$
\begin{equation*}
\sum_{t=1}^{T}\left[ y_{t}(N)-y_{t}^{\ast }(N)\right] d_{t}^{i}=\sum
_{\substack{ 1\leq t\leq T  \\ d_{t}^{N}=0}}\left[ y_{t}(N)-y_{t}^{\ast }(N)%
\right] d_{t}^{i}+\sum_{\substack{ 1\leq t\leq T  \\ d_{t}^{N}>0}}\left[
y_{t}(N)-y_{t}^{\ast }(N)\right] d_{t}^{i}=0.
\end{equation*}
\end{proof}

Every cost allocation $o=(o_{i})_{i\in N}$ of a PI-game defined in
this way is what we call the \emph{Owen point} for $(N,c).$
Obviously,\emph{\ }the Owen point is a core-allocation. The Owen
point, for each player $i$, can be interpreted as the cost he/she
has to pay when producing at the minimum operation cost. Hence, we
propose the Owen point as an alternative value for PI-games. We note
in passing that if the matrices $P$, $H$ and $B$ are integers then
the core-allocation given by the Owen point is also in integer
values.

The Owen point for the situation given in Example \ref{cf} is obtained by

\begin{equation*}
\begin{array}{l}
o_{1}=p_{2}^{N}d_{2}^{1}+d_{3}^{1}(h_{23}^{N}+p_{2}^{N})+d_{1}^{1}(b_{21}^{N}+p_{2}^{N})=7,
\\
o_{2}=p_{2}^{N}d_{2}^{2}+d_{3}^{2}(h_{23}^{N}+p_{2}^{N})+d_{1}^{2}(b_{21}^{N}+p_{2}^{N})=14,
\\
o_{3}=p_{2}^{N}d_{2}^{3}+d_{3}^{3}(h_{23}^{N}+p_{2}^{N})+d_{1}^{3}(b_{21}^{N}+p_{2}^{N})=8.%
\end{array}%
\end{equation*}

\noindent Next theorem shows that there is a pmas that realizes the Owen
point.

\begin{theorem}
Let $(N,D,H,B,P)$ be a PI-situation with $D$ being an integer matrix, and $%
(N,c)$ the corresponding PI-game. The Owen point can be reached through a
pmas.
\end{theorem}
\begin{proof}
Define for all $i\in S,S\subseteq N$ and $S\neq \varnothing ,$

\begin{equation*}
y_{i}^{S}:=\sum_{t=1}^{T}y_{t}^{\ast }(S)d_{t}^{i}.
\end{equation*}

Then for all $S\subseteq N,S\neq \varnothing $

\begin{equation*}
y^{S}(S)=\sum_{i\in S}\left( \sum_{t=1}^{T}y_{t}^{\ast }(S)d_{t}^{i}\right)
=\sum_{t=1}^{T}y_{t}^{\ast }(S)d_{t}^{S}=c(S),
\end{equation*}%
and for all $S\subseteq R\subseteq N,S,R\neq \varnothing $ and for all $i\in
S,$

\begin{equation*}
y_{i}^{S}=\sum_{t=1}^{T}y_{t}^{\ast }(S)d_{t}^{i}\geq
\sum_{t=1}^{T}y_{t}^{\ast }(R)d_{t}^{i}=y_{i}^{R},
\end{equation*}%
since $y_{t}^{\ast }(S)\geq y_{t}^{\ast }(R)$ (a consequence of $%
F(DLPI(R))\subseteq F(DLPI(S))$ and Lemma \ref{facti} extended to
the coalition $S\subseteq N).$

Finally, we see that $y_{i}^{N}=o_{i}$ for all $i\in N.$ So, the Owen point $%
o$ can be reached through the pmas $\left( y^{S}\right) _{\varnothing \neq
S\subseteq N}.$
\end{proof}

Existence of pmas for the entire class of Production-Inventory games
is important by itself but in addition has an interesting
consequence. It allows us to prove that PI-games are strictly
included in the class of totally balanced games  since in general
there are totally balanced games without pmas.

It is also possible to prove the existence of pmas in PI-games using
the following analysis: consider a situation with $N$ players and
$T$ time periods, but only $1$ out of the $|N|\times|T|$ demands is
positive. The player with this demand pays the costs, which
obviously decreases if the coalition he belongs to increases. This
is a natural pmas (this player is some sort of a veto-player).
Extending this to a general setting by adding $|N| \times |T|$ of
these games provides a pmas in general.

Another observation, from the proof of the above theorem, is that a
pmas for any PI-game can be built just taking the Owen point of each
subgame and gathering all of them as a collection of vectors.

\noindent The Owen point for the game given in Example \ref{cf} can be
reached through the pmas
\begin{equation*}
\left( \left( 7\right) ^{\{1\}},\left( 28\right) ^{\{2\}},\left( 8\right)
^{\{3\}},\left( 7,14\right) ^{\{1,2\}},\left( 7,8\right) ^{\{1,3\}},\left(
14,8\right) ^{\{2,3\}},\left( 7,14,8\right) ^{\{1,2,3\}}\right) .
\end{equation*}

In that example the core is a singleton: the Owen point. This is not
a general property for PI-games. Next theorem provides a necessary
and sufficient condition for the core of a PI-game to be a
singleton. The following concept is required.

\begin{definition}
Let $(N,D,H,B,P)$ be a PI-situation with $D$ being an integer
matrix. We say that $i\in N$ is an \emph{essential player} for
$(N,D,H,B,P)$ if there exists at least one period $t\in \{1,..,T\}$
with $d_{t}^{N\backslash \{i\}}>0$ such that $y_{t}^{\ast
}(N)<y_{t}^{\ast }(N\backslash \{i\}).$
\end{definition}

The reader may notice that an essential player is the one for which at least
one period exists, in which it is needed by the rest of players in order to
produce at a minimum cost a certain demand. On the other hand, an \emph{%
inessential player }$i$ is the one which is unnecessary by the grand
coalition to operate at minimum cost.

We denote by $\mathcal{E}$ the set of essential players and by
$N\backslash \mathcal{E}$ its complementary set. In the following we
prove that the core of a PI-game shrinks to the Owen point just only
when all players are inessential for the PI-situation.

\begin{theorem}
Let $(N,D,H,B,P)$ be a PI-situation with $D$ being an integer matrix and $%
(N,c) $ the corresponding PI-game. Then, $Core(N,c)=\{o\}$ if and only if $%
\mathcal{E=\varnothing }.$
\end{theorem}
\begin{proof}
\emph{(Only if)} Take $(N,D,H,B,P)$ a PI-situation with $D$ being an integer
matrix and all players are inessential for it. Then, for all $i\in N,$ $%
y_{t}^{\ast }(N\backslash \{i\})=y_{t}^{\ast }(N)$ for all $t\in \{1,...,T\}$
with $d_{t}^{N\backslash \{i\}}>0$, and $y_{t}^{\ast }(N\backslash
\{i\})\geq y_{t}^{\ast }(N)$ for all $t\in \{1,...,T\}$ with $%
d_{t}^{N\backslash \{i\}}=0.$ Hence, $c(N\backslash \{i\})=o(N\backslash
\{i\}).$

Take $x\in Core(N,c)$ and recall that $o\in Core(N,c)$. Then, for
all $i\in N,c(N)-c(N\backslash \{i\})=o_{i}\leq x_{i}$ and
$\sum_{i=1}^{n}x_{i}=\sum_{i=1}^{n}o_{i}.$
Hence, $x_{i}=o_{i}$ for all $i\in N$. So we can conclude that $%
Core(N,c)=\{o\}.$

\emph{(If)} Suppose that $\mathcal{E\neq \varnothing }.$ Take $j\in \mathcal{%
E}$, then there exists $t^{\ast }\in \{1,...,T\}$ with $d_{t^*}^{N\backslash
\{j\}}>0$ such that $y_{t^{\ast }}^{\ast }(N)<y_{t^{\ast }}^{\ast
}(N\backslash \{j\}).$ For any period $t$ satisfying the above condition,
there exists $i\in N\backslash \{j\}$ such that $d_{t}^{i}>0.$ It can be
checked that for all $R\subset N\backslash \{j\},i\in R$ it holds $%
y_{t}^{\ast }(N)<y_{t}^{\ast }(N\backslash \{j\})\leq y_{t}^{\ast }(R)$.
Hence $o(R)<c(R).$

Define $\Delta :=\{R\subset N\setminus \{j\}: i\in R \},\; 0<\alpha :=%
\underset{R\in \Delta }{\min }\{c(R)-o(R)\},$ and for all $k\in N$%
\begin{equation}
o_{k}^{\ast }=\left\{
\begin{array}{cc}
o_{k} & k\in N\backslash \{i,j\}, \\
o_{i}+\alpha & k=i, \\
o_{j}-\alpha & k=j.%
\end{array}%
\right.  \label{eq:newop}
\end{equation}

It can be easily checked that $o^{\ast }\in Core(N,c).$ Hence, we can
conclude that $Core(N,c)\neq \{o\}.$
\end{proof}

The above proof sheds light onto the structure of the core. If there
exists at least one essential player, the rest of players compensate
him for cooperation by reducing the cost when producing at minimum
production cost according to (\ref{eq:newop}). On the contrary, if
no player is essential the unique core-allocation is the one given
by the cost generated when producing at minimum production cost.

We would like to conclude this section positioning the Owen point in
comparison with other well-known allocations in cooperative game
theory. Although due to space requirements we do not include
examples in the paper, the interested reader can find examples
supporting our claims in the working paper by Guardiola et al.
(2004). First of all, in the class of PI-games the Shapley value
(Shapley, 1953) is not, in general, a core allocation. Moreover,
even in those cases when Shapley value belongs to the core it does
not have to coincide with the Owen point. Finally, we point out that
the Owen point does not coincide either with the nucleolus
(Schmeidler, 1969).

\section{\label{COR}\textbf{Concluding Remarks}}

In this paper we have presented a model of cooperation among several
firms arising from dynamic production-inventory situations with
discrete demand and finite planning horizon. In this model any group
of firms can agree to cooperate because fair allocations of
operation costs which are stable always exist. In this sense none of
the firms would have an incentive to leave the group. From among all
the above fair allocations, we propose the Owen point: the
allocation in which every firm has to pay the minimum cost of
operation. This allocation is especially appealing since it can be
calculated in polynomial time and, moreover, it can be reached
through a
pmas. 

We note in passing that incorporating selling prices to the considered model
would have not modified the structure of the problem, so that all the
conclusions would have been the same.

We would like to finish these remarks mentioning some additional topics for
further research on the cooperation model considered in this paper: (1) to
find out descriptions of the structure of the core, (2) axiomatic
characterizations\ of\ the\ Owen\ point, (3) alternative allocation schemes
and (4) other forms of cooperation and /or competition, as well as models
with concave cost functions.



\clearpage

\begin{thebibliography}{9}


\bibitem{Ze91} {Anupindi, R., Bassok, Y., Zemel, E., 2001. A
general framework for the study of decentralized distribution
systems. Manufacturing \& Service Operations Management. 3,
349-368.}

\bibitem{B63} { Bondareva, O.N., 1963. Some applications of
linear programming methods to the theory of cooperative games. Prob.
Kibernety 10, 119-139. In Russian.}

\bibitem{BHH01} { Borm, P.E.M., Hamers, H., Hendrickx, R., 2001.
 Operations Research Games: A Survey. TOP 9, 139-216.}


\bibitem{DIN99} { Deng, X., Ibaraki, T., Nagamochi, H., 1999.
Algorithmic aspect of the core of combinatorial optimization games.
Math.Oper. Res. 24, 751-766.}

\bibitem{DINZ00} { Deng, X., Ibaraki, T., Nagamochi, H., Zang, W., 2000.
 Totally balanced combinatorial optimization
games. Math. Programming 87, 441-452.}

\bibitem{DS84} { Dubey, P., Shapley, L.S., 1984. Totally
balanced games arising from controlled programming problems. Math.
Programming 29, 245-267.}

\bibitem{E79} { Eppen, G.D., 1979. Effect of Centralization
on Expected Cost in a Multi-location Newsboy Problem. Manage. Sci.
25, 498-501.}

\bibitem{FK00} { Faigle, U., Kern, W., 2000. On the core
of ordered submodular cost games. Math. Programming 87, 483-499.}

\bibitem{GN72} { Garfinkel, R.S., Nemhauser, G.L., 1972. Integer
Programming. Wiley, New York.}

\bibitem{GPRE00} { Gellekom, J.R.G., Potters, J.A.M., Reijnierse, J.H., Engel,
M.C., Tijs, S.H., 2000. Characterization of the Owen Set of Linear
Production Processes. Games Econ. Behav. 32, 139-156.}

\bibitem{GG91} { Gerchak, Y., Gupta, D., 1991. On
Apportioning Costs to Customers in Centralized Continuous Review
Inventory Systems. J.Oper. Manage. 10, 546-551.}

\bibitem{Gra86} { Granot, D. 1986. A generalized linear
production model: A unified model. Math. Programming 34, 212-222.}


\bibitem{GMJ04a} { Guardiola, L.A., Meca, A., Puerto, J., 2004.
Production-Inventory games: a new class of totally balanced
combinatorial optimization games. Trabajos de I+D I-2004-8, CIO
Universidad Miguel Hern\'{a}ndez, Elche.}


\bibitem{HD96} { Hartman, B.C., Dror, M., 1996. Cost
allocation in continuous review inventory models. Naval Res. Logist.
43, 549-561.}

 \bibitem{HD03} { Hartman, B.C., Dror, M., 2003.
Optimizing centralized inventory operations in a cooperative game
theory seting. IIE Trans. 35, 243-257.}

\bibitem{HD05} { Hartman, B.C., Dror, M., 2005.
Allocation of gains from inventory centralization in newsvendor
environments. IIE Trans. 37, 93-107.}

\bibitem{HDS00} { Hartman, B.C., Dror. M., Shaked, M., 2000.
Cores of inventory centralization games. Games Econ. Behav. 31,
26-49.}

\bibitem{JM74} { Johnson, L.A., Montgomery, D.C., 1974.
Operations Research in Production Planning, Scheduling, and
Inventory Control. John Wiley \& Sons.}

 \bibitem{KP89} { Kohli, R., Park, H., 1989. A
cooperative game model of quantity discounts. Manage. Sci. 35,
693-707.}

\bibitem{M04} { Meca, A., 2006. A core-allocation family for
generalized holding cost games. Math. Methods Oper. Res. (to
appear).}

\bibitem{MTGB04} { Meca, A., Timmer, J., Garc\'{i}a-Jurado, I., Borm,
P.E.M., 2004. Inventory Games. European J. Oper. Res. 156, 127-139.}

\bibitem{MGB03} { Meca, A., Garc\'{i}a-Jurado, I., Borm, P.E.M., 2003.
Cooperation and competition in Inventory Games. Math. Methods Oper.
Res. 57, 481-493.}

\bibitem{M03} { Minner, S., 2006. Bargaining for cooperative
economic ordering. Decis. Support Syst. (in press).}

\bibitem{MSS02} { M\"{u}ller, A., Scarsini, M., Shaked, M., 2002.
The Newsvendor Game has a Nonempty Core. Games Econ. Behav. 38,
118-126.}

\bibitem{R93} { Robinson, L.W., 1993. Comment on \textquotedblleft On
Apportioning Costs to Customers in Centralized Continuous Review Inventory
Systems,\textquotedblright\ by Gerchak and Gupta, J. Oper. Manage. 11%
, 99-102.}

\bibitem{SCH79} { Schmeidler, D., 1969. The
Nucleolus of a Characteristic Funtion Game. SIAM J. Appl. Math. 17,
1163-1170.} 


\bibitem{SH53} { Shapley, L.S., 1953. A Value for n-Person
Games. In: Kuhn H, Tucker AW (eds.) Contributions to the Theory of
Games II. Princeton University Press, pp. 307-317.}

\bibitem{SH67} { Shapley, L.S., 1967. On Balanced Sets and
Cores. Naval Res. Logist. 14, 453-460.}

\bibitem{SH71} { Shapley, L.S., 1971. Cores of Convex
Games. Int. J. Game Theory 1, 11-26.}

\bibitem{SS69} { Shapley, L.S., Shubik, M., 1969. On
market games. J. Econ. Theory 1, 9-25.}

\bibitem{SS72} { Shapley, L.S., Shubik, M., 1972. The
assignment game. Int. J. Game Theory. 1, 111-130.}

\bibitem{SFW01} { Slikker, M., Fransoo, J., Wouters, M., 2005.
Cooperation between multiple news-vendors with transshipments.
European J. Oper. Res. 167, 370-380.}

\bibitem{S90} { Sprumont, Y., 1990. Population Monotonic
Allocation Schemes for Cooperative Games with Transferable Utility.
Games Econ. Behav. 2, 378-394.}

\bibitem{Tam92} { Tamir, A., 1992. On the Core of Cost
Allocation Games Defined on Location Problems. Transp. Sci. 27,
81-86.}


\bibitem{TML00} { Tijs, S.H., Meca, A., L\'{o}pez, M.A., 2005.
Benefit sharing in holding situations. European J. Oper. Res. 162,
251-269.} 

\end{thebibliography}
\end{document}